\documentclass[pra,nofootinbib,floatfix,10pt,twocolumn]{revtex4-2}

\usepackage{amsmath}
\usepackage{amssymb}
\usepackage{wasysym}
\usepackage{graphicx}
\usepackage{color,soul}
\usepackage{physics}
\usepackage{siunitx}
\usepackage{dsfont}
\usepackage{float}
\usepackage[english]{babel}
\usepackage{blindtext}
\usepackage[english,nomargin,inline,marginclue,draft]{fixme}
\pdfpageheight\paperheight
\pdfpagewidth\paperwidth
\newcommand{\lyxdot}{.}
\usepackage[colorlinks,linkcolor=blue,anchorcolor=blue,citecolor=blue,urlcolor=blue]{hyperref}
            
\fxusetheme{colorsig}
\FXRegisterAuthor{cg}{acg}{CG}  
\FXRegisterAuthor{th}{ath}{\color{blue}TH}  
\FXRegisterAuthor{ib}{aib}{\color{red}IB} 
\FXRegisterAuthor{sh}{ash}{\color{cyan}SH} 
\FXRegisterAuthor{db}{adb}{\color{green}DB} % now I can ..
\FXRegisterAuthor{ps}{aps}{PS}
\makeatletter
\renewcommand*\FXLayoutInline[3]{%
  {\@fxuseface{inline}\ignorespaces{\color{fx#1}[#3: #2]}}}
\makeatother

\long\def\symbolfootnote[#1]#2{\begingroup%
\def\thefootnote{\fnsymbol{footnote}}\footnotetext[#1]{#2}\endgroup}

\def\nobreakbefore{%
  \relax\ifvmode\else
    \ifhmode
      \ifdim\lastskip > 0pt\relax
        \unskip\nobreakspace
      \else % added to put a ~if no space was typed. (Unclear why it sometimes worked before )
        \nobreakspace
      \fi
    \fi
  \fi
}
\let\oldcite\cite
\renewcommand\cite{\nobreakbefore\oldcite}

\begin{document}
\title{Enhanced metrology at the critical point of a many-body Rydberg atomic system}

\author{Dong-Sheng Ding$^{1,2,3,\textcolor{blue}{\ast},\textcolor{blue}{\dag}}$}
\author{Zong-Kai Liu$^{1,2,3,\textcolor{blue}{\ast}}$}
\author{Bao-Sen Shi$^{1,2,3,\textcolor{blue}{\ddagger}}$}
\author{Guang-Can Guo$^{1,2,3}$}
\author{Klaus Mølmer$^{4,\textcolor{blue}{\mathsection}}$}
\author{Charles S. Adams$^{5,\textcolor{blue}{\mathparagraph}}$}

\affiliation{$^1$CAS Key Laboratory of Quantum Information, University of Science and Technology of China, Hefei 230026, China}
\affiliation{$^2$CAS Center for Excellence in Quantum Information and Quantum Physics, University of Science and Technology of China, Hefei 230026, China}
\affiliation{$^3$Hefei National Laboratory, University of Science and Technology of China, Hefei 230088, China}
\affiliation{$^4$Aarhus Institute of Advanced Studies, Aarhus University, H{\o}egh-Guldbergs
Gade 6B, DK-8000 Aarhus C, Denmark\\
Center for Complex Quantum Systems, Department of Physics and Astronomy,
Aarhus University, Ny Munkegade 120, DK-8000 Aarhus C, Denmark}
\affiliation{$^5$Department of Physics, Durham University, South Road, Durham DH1 3LE, United Kingdom}
\date{\today}

\symbolfootnote[1]{These authors contributed equally to this work.}
\symbolfootnote[2]{dds@ustc.edu.cn}
\symbolfootnote[3]{drshi@ustc.edu.cn}
\symbolfootnote[4]{moelmer@phys.au.dk}
\symbolfootnote[5]{c.s.adams@durham.ac.uk}

\maketitle

\textbf{The spectral properties of an interacting many-body system may display critical character and have potential applications in precision metrology. Here, we demonstrate such many-body enhanced metrology for microwave (MW) electric fields in a non-equilibrium Rydberg atomic gas. Near criticality the high sensitivity of Rydberg atoms to external MW electric fields, combined with many-body enhancement induces significant changes in the optical transmission. We quantify this behavior using the Fisher information. For continuous optical transmission at the critical point, the Fisher information is three orders of magnitude larger than in independent particle systems, the 
measured data provides an equivalent sensitivity of 49 nV/cm/$\textrm{Hz}^{1/2}$.
The reported results constitute a milestone towards the application of many-body effects in precision metrology.}

\section*{Introduction}
Ensembles of well-controlled neutral atoms are ideal systems to explore many-body physics \cite{greiner2002quantum,bernien2017probing,martin2013quantum,colombo2021time}. In particular, the controllable interactions among highly-excited Rydberg atoms hold promise for studies of quantum information and many-body physics \citep{lukin2001dipole,saffman2010quantum,carr2013nonequilibrium}. Benefiting from the large interaction volume of Rydberg atoms, a small change in the Rydberg state population can induce a global macroscopic phase transition between non-interacting (NI) and interacting (I) phases \citep{ding2019Phase}. Laser-induced density-dependent energy shifts of Rydberg states offer a convenient platform to directly observe non-equilibrium phase transitions and bistability \citep{carr2013nonequilibrium,malossi2014full,de2016intrinsic,vsibalic2016driven,wade2018terahertz}, and to study dynamical analogues of forest fire \citep{ding2019Phase} and epidemic spreading \citep{wintermantel2020epidemic,ding2021}.
In contrast to other optically bistable systems \citep{gibbs1976differential,wang2001bistability,wang2001enhanced,pickup2018optical,hehlen1994cooperative}, Rydberg ensemble experiments can be performed  without the need of optical cavity feedback and cryogenic temperatures. 

Exploring the non-equilibrium dynamics of the Rydberg system under external fields is intriguing. The emergent thermodynamic and spectroscopic properties of a many-body system of interacting Rydberg atoms present open questions both in theory \citep{lee2012collective,marcuzzi2014universal,weimer2015variational,vsibalic2016driven,levi2016quantum} and experiments \citep{carr2013nonequilibrium,ding2019Phase}. Due to the large dipole moment, the Rydberg atoms are highly sensitive to system noise and external electric fields \citep{fan2015atom,sedlacek2012microwave,facon2016sensitive,cox2018quantum,jing2020atomic,Liu2022Deep}. Most dramatically, the macroscopic change in the optical response near a critical point \citep{ding2019Phase,wade2018terahertz} presents a resource for increased metrological sensitivity \citep{gammelmark2011phase, macieszczak2016dynamical, PhysRevA.96.013817,raghunandan2018high,PhysRevLett.124.120504, PhysRevLett.126.010502,PhysRevLett.126.200501,Theodoros2021Criticality,garbe2021critical,liu2021experimental,PhysRevA.78.042105}. Accompanying the divergent susceptibility near the critical point, optical probing of the system is highly sensitive to small variations of physical parameters. Critical systems may thus display sensing errors with a generic scaling $\sim1/\sqrt{N^{\gamma}t^{\lambda}}$ with $\gamma,\lambda>1$ \citep{Theodoros2021Criticality,rossini2020dynamic,pelissetto2018dynamic}, where $N$ is the number of atoms and $t$ is the measurement time.

In this article, we demonstrate how Rydberg criticality provides a method for high-sensitivity probing of external parameters. We exploit the extreme sensitivity of the optical transmission at the critical point to probe external MW fields. Due to the critical slowing down near phase transitions, we need to take into account how the system dynamics do not adiabatically follow the stationary state, but rather smooths the system response. This leads to a non-integer power dependence of the Fisher Information (FI) on the duration of the detuning scans. The behavior around criticality is observed to enhance the FI by a factor of up to more than $10^3$ compared to a non-interacting ensemble.

\begin{figure*}[t]
\includegraphics[width=2\columnwidth]{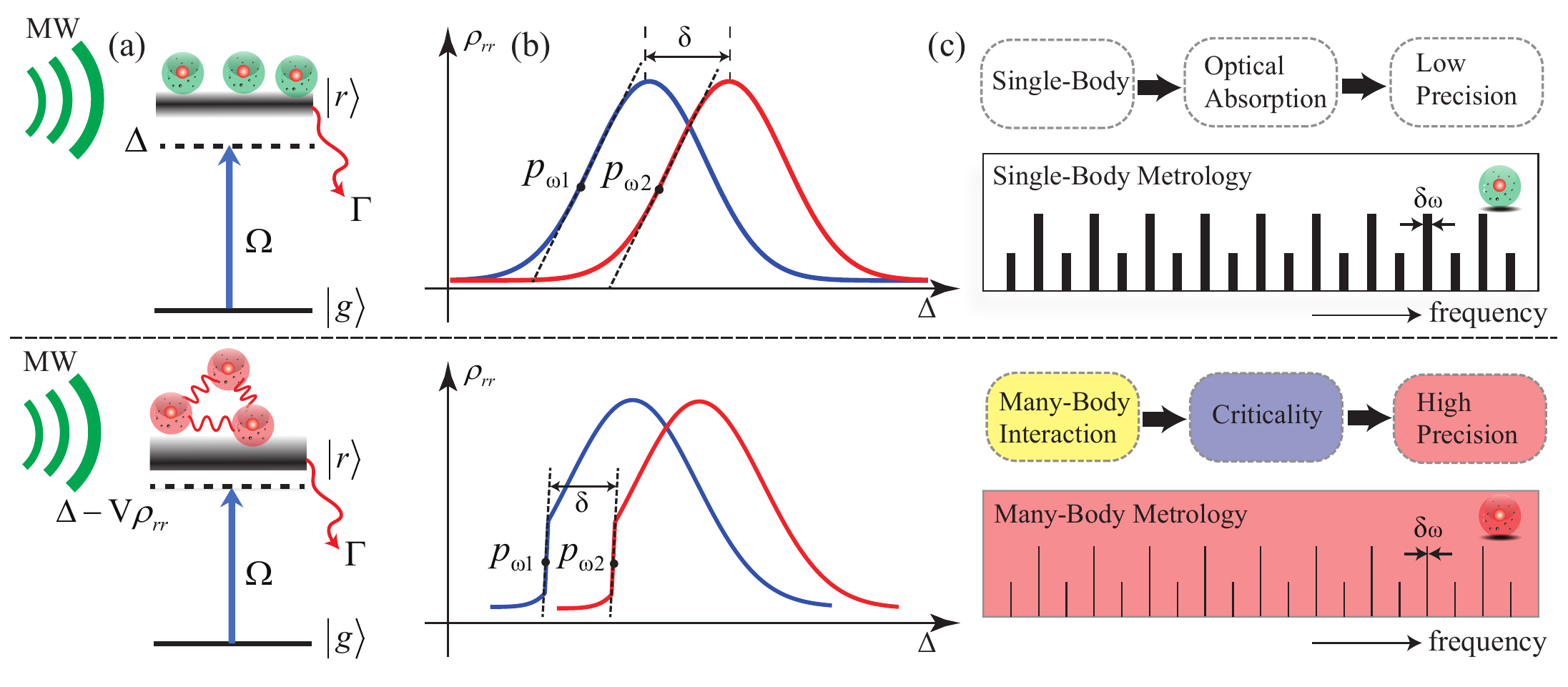}\caption{\textbf{The principle of single-body [top row] (many-body [bottom row]) Rydberg metrology} (a) Energy diagram for a two-level atom model, showing the ground state $\left| g \right\rangle$ and  Rydberg state $\left| r \right\rangle$ with spontaneous radiation rate, $\Gamma$. The atoms are driven from the ground state to the Rydberg state by a laser with Rabi frequency, $\Omega$ and detuning, $\Delta$. They are also exposed to a MW field with electric field component $E_{mw}$. In the many-body case, the Rydberg resonance is modified by the many-body interaction strength, $V = {C_6}/{r^6} $ [where $C_6$ is the van der Waals coefficient and $r$ is the distance between Rydberg atoms], and the population of the Rydberg atoms $\rho_{rr}$; see more details in the Method Sections.
(b) The blue and red curves represent the spectrum with and without external MW field, which induces a shift $\delta$. The measurement sensitivity is highest when the derivative $\mathrm{d}\rho_{rr}/d\Delta$ is maximal, as indicated by the points $p_{w1}$ and $p_{w2}$. The steeper slope near the critical point in the many-body case [bottom row] results in enhanced measurement sensitivity.  (c) The many-body advantage corresponds to a metrological ruler with thinner tick marks $\delta \omega$ than in the single-body case. The transmission spectra are shifted by an external electrical field forming a ruler with unfixed ticks.}

\label{setup}
\end{figure*}

\section*{Results}
\textbf{Many-body metrology model.}
We consider a model of $N$ interacting two-level atoms with a ground state $\left| g \right\rangle$ and a Rydberg state $\left| r \right\rangle$ (with decay rate $\Gamma$) [Fig.~\ref{setup}(a)]. A laser couples these atoms with Rabi frequency $\Omega$ and detuning $\Delta$. We derive the Rydberg population $\rho_{rr}$ via mean-field approximation (i.e., $\Delta \rightarrow \Delta - V\rho_{rr}$, where $V$ is the average many-body interaction strength from dipole interaction or ions collisions), $\delta$ is the external field induced-frequency shift on Rydberg state $\left| r \right\rangle$, as mentioned in Method Sections,
\begin{equation}
\rho_{\mathrm{rr}}=\frac{\Omega^{2}}{4\left(\Delta-V\rho_{\mathrm{rr}}\right)^{2}+2\Omega^{2}+\Gamma^{2}}.
\end{equation}
Due to the interaction, the spectrum has a population-dependent shift $V\rho_{rr}$, thus inducing a steep edge of $\rho_{rr}$ with a maximum derivative
\begin{equation}
\left.\frac{\mathrm{d}\rho_{rr}}{\mathrm{d}\Delta}\right|_{\Delta=\Delta_{c}}=\frac{1}{V+\sqrt{(\Gamma^{2}+2\Omega^{2})/3\rho_{rr}^{2}}},
\end{equation}
where $\Delta_{c}$ corresponds to the detuning at which the derivative gets its maximum. We note that the $\mathrm{d}\rho_{rr}/\mathrm{d}\Delta$ is increased due to the interaction strength $V$ (here $V<0$); more details can be found in Method Sections. The derivative $\mathrm{d}\rho_{rr}/\mathrm{d}\Delta$ diverges at the system's critical point \cite{ding2019Phase}, exhibiting a method of high-precision measurement \cite{gammelmark2011phase, macieszczak2016dynamical, PhysRevA.96.013817,raghunandan2018high,PhysRevLett.124.120504, PhysRevLett.126.010502,PhysRevLett.126.200501}. A measurement is realized by detecting the transmission of an optical probe field. When applying external fields (such as the electric component of external microwave MW fields), the measurement precision is limited by the maximum slope of the Rydberg resonance. This is indicated by the points $p_{w1}$ ($p_{w2}$) in [Fig.~\ref{setup}(b)]. Compared to the non-interacting case [top row], the slope in the vicinity of the critical point [bottom row] is significantly enhanced. For metrology applications, the sensitivity of many-body case is enhanced by a ratio,
\begin{equation}
\text{\ensuremath{\beta}=}\left.\frac{\mathrm{d}\rho_{rr}}{\mathrm{d}\Delta}\right|_{V\neq0}/\left.\frac{\mathrm{d}\rho_{rr}}{\mathrm{d}\Delta}\right|_{V=0}.
\end{equation}
In Fig.~\ref{setup}(c) we illustrate how this many-body enhancement is like having a new ruler with much finer markings.

\begin{figure}
\centering
\includegraphics[width=1\columnwidth]{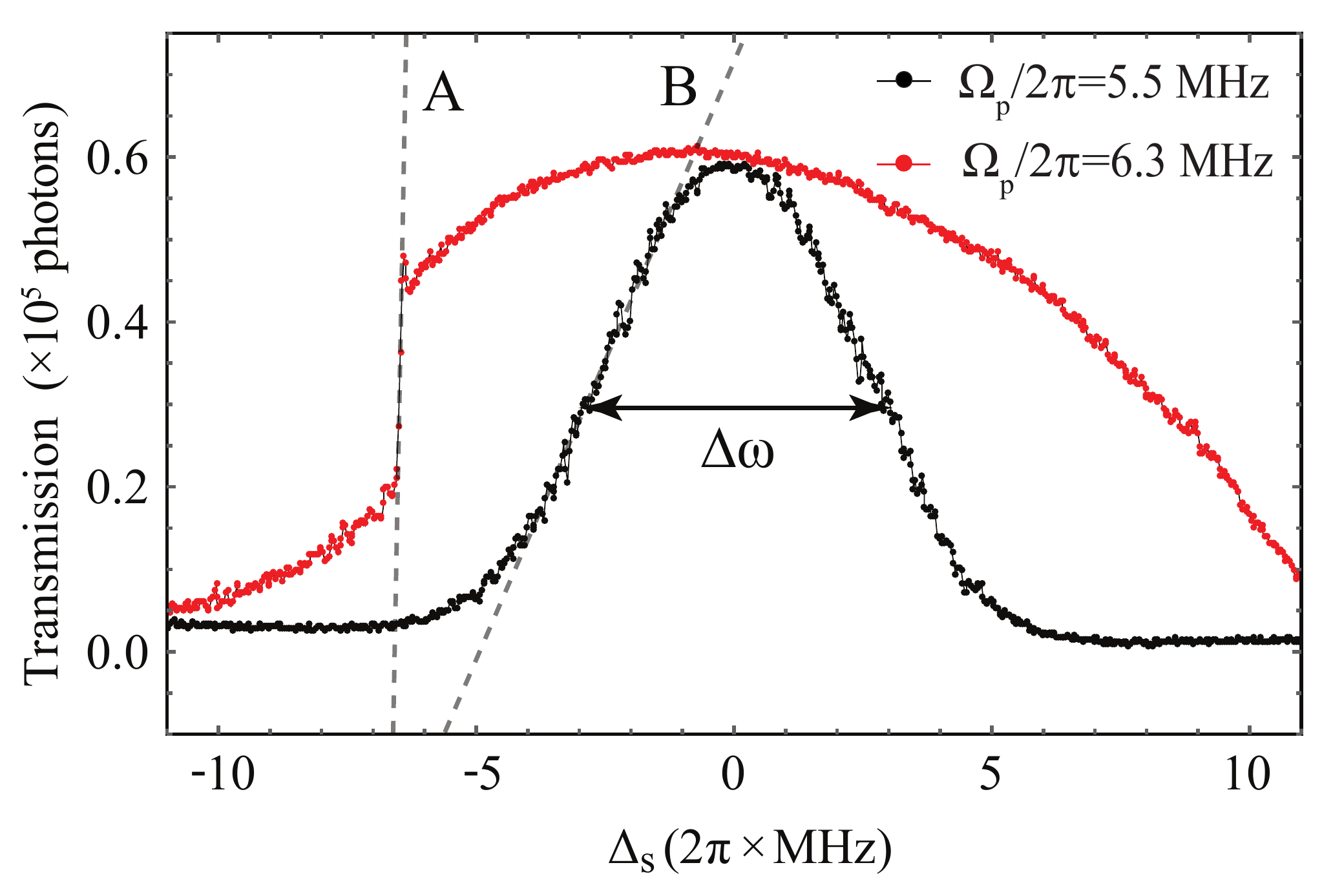}\\
\caption{\textbf{Optical transmission spectra with and without phase transition.} The transmission spectra with (red) and without (black) phase transition. The dashed lines A and B show the maximum slopes near half transmission. $\Delta{\omega}/2\pi\sim$ 6 MHz shows the bandwidth of the transmission spectrum without interactions. The photon counts are given for a measurement time $t$ = 20 $\mu$s for $2\pi\times$0.036 MHz detuning interval.}\label{slope}
\end{figure}

The measurement sensitivity is determined by the variation of the transmission signal around the critical point and the photon counting noise in the measurement record. By exploring the linear slope of transmission in a narrow interval around the critical point the sensitivity can be expressed in terms of the Fisher Information
\cite{RevModPhys.90.035005,braunstein1996generalized},
\begin{equation}
F(\Delta)=\frac{\overline{\mu}'(\Delta)^{2}}{\overline{Var(\mu)}}
\label{eq:SNR}
\end{equation}
where $\Delta$ is the parameter that we want to determine, $\overline{ \mu }$ represents the mean value of the difference in photon numbers accumulated in fixed time intervals by a differential detector exposed to a reference beam and a beam passing through the atomic cloud. $Var(\mu)$ is the variance of the differential signal, i.e  the sum of the variances of the two  separate and indendent counting signals. Note that Eq.(4) expresses the usual signal-to-noise ratio, and the Cram\'{e}r-Rao bound,
\begin{equation}
\delta\Delta\geq\frac{1}{\sqrt{ F(\Delta)}}.
\label{eq:CR}
\end{equation}
yields the usual estimation error for the counting signal.

It is important to emphasize that the Fisher Information refers to the actual counting signals. While a detector may output a photon rate in counts per second, we must independently assess or estimate the variance of the signal for the time duration of the measurement. We explore experiments with different duration and we shall thus present the actual counts in given time intervals and their variance to obtain the proper assessment of the metrological sensitivity. We shall also observe that during a frequency scan in finite time, the non-interacting (interacting) atomic system does (does not) attain its stationary state. This leads to a linear (non-linear) dependence of the Fisher information on the measurement time.  

\begin{figure*}
\centering
\includegraphics[width=2\columnwidth]{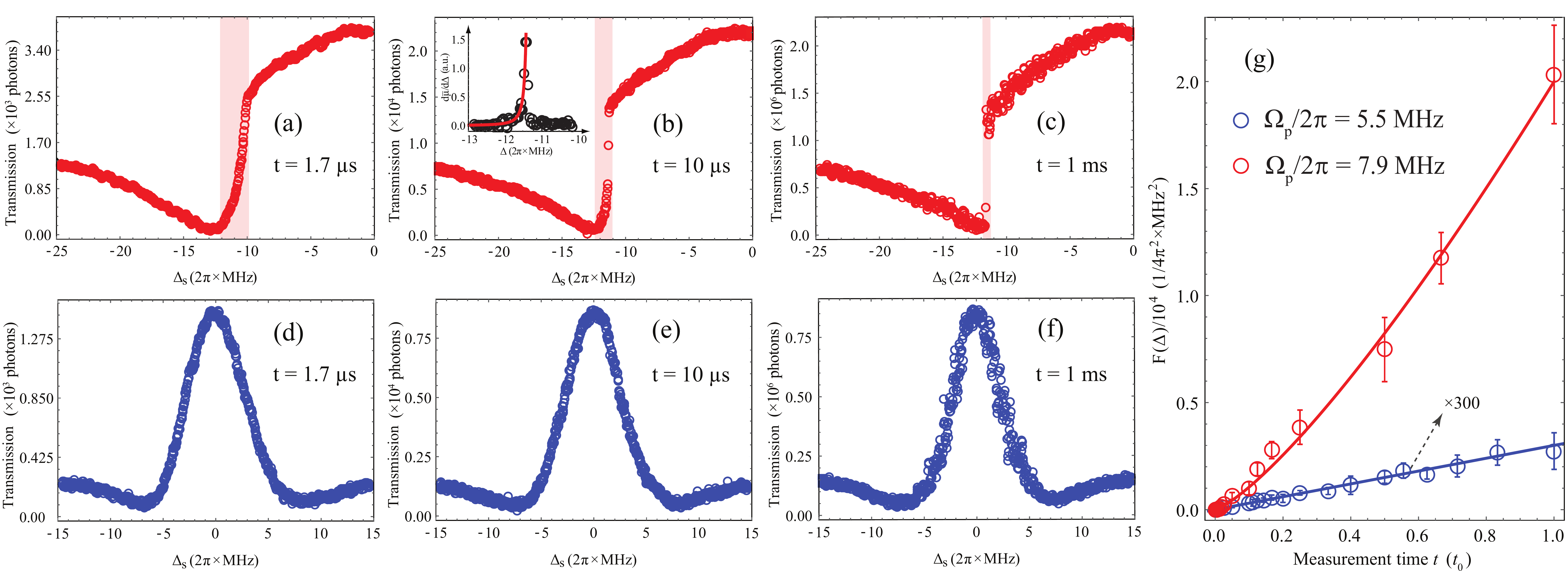}\\
\caption{\textbf{Transmission spectra and the associated  Fisher information.} Panels (a)-(c) show the transmission spectra across the phase transition (shaded regions), obtained in a total measurement time of $t=1.7$ $ \mu$s, $t=10$ $\mu$s, and $t=1$ ms. The inset in (b) shows the corresponding derivative $d\overline{\mu}/d\Delta_{s}$. Panels (d)-(f) show the trivial spectra (without phase transition) for the same measurement times $t=1.7$ $\mu$s, $t$=10 $\mu$s, and $t$= 1 ms. Panel (g) shows the Fisher information associated with the determination of the steepest point on the transmission curves for different values of the total measurement time $t$ (note that the Fisher Information for the non-interacting case is magnified manually by a factor 300. The red and blue curves are fitted by the function $F=A(t/t_{0})^{\lambda}$, where the fit parameters are given in the main text. In this process, the red data in (g) is obtained from the maximum of \textbf{$F(\Delta)|_{\Delta_{s}=\Delta_{c}}=(\mathrm{d}\overline{\mu}/\mathrm{d} \Delta_s)^{2}/Var(\mu)$}, while for the blue data in (g), the FI at the critical point are obtained by considering 30 data points around $\Delta_{c}$ in (d)-(f) to reduce the fluctuations from the instability of the laser power and the cell temperature. The error bars determined in (g) are statistics from the three repeated experiments.} \label{Fishervstime}
\end{figure*}

\textbf{Measured derivative and Fisher information.} 
For the experiment, we employ two-photon excitation with a probe beam and a coupling laser beam with
Rabi frequencies (detuning) $\Omega_{p}$ ($\Delta_{s}$) and $\Omega_{c}$ ($\Delta_{c}$). We measure the transmission spectra, as shown in Fig.~\ref{slope}(a). The transmission depends on $\Omega_{p}$, and we can prepare the system with and without phase transition $\Omega_{p}<\Omega_{p,c}$ and $\Omega_{p}>\Omega_{p,c}$, where $\Omega_{p,c}\text{/}2\pi=5.6$ MHz is the threshold Rabi frequency of the probe field. Our system displays a second order dynamical phase transition between two stationary states with different excitation densities \citep{marcuzzi2014universal}. The two spectra display the main distinct character of the transmission of a non-interacting system (the black curve in Fig.~\ref{slope}(a)), and an interacting many-body system (the red curve in Fig.~\ref{slope}(a)). The derivative $\mathrm{d}\mu/\mathrm{d}\Delta_{s}$ of the transmission is very large  at the phase transition point $\Delta_c$  of the red curve in Fig.~\ref{slope}(a), while it explores a weaker, finite maximum near half maximum of the black black curve in Fig.~\ref{slope}(a).

\begin{figure*}[t]
\includegraphics[width=2\columnwidth]{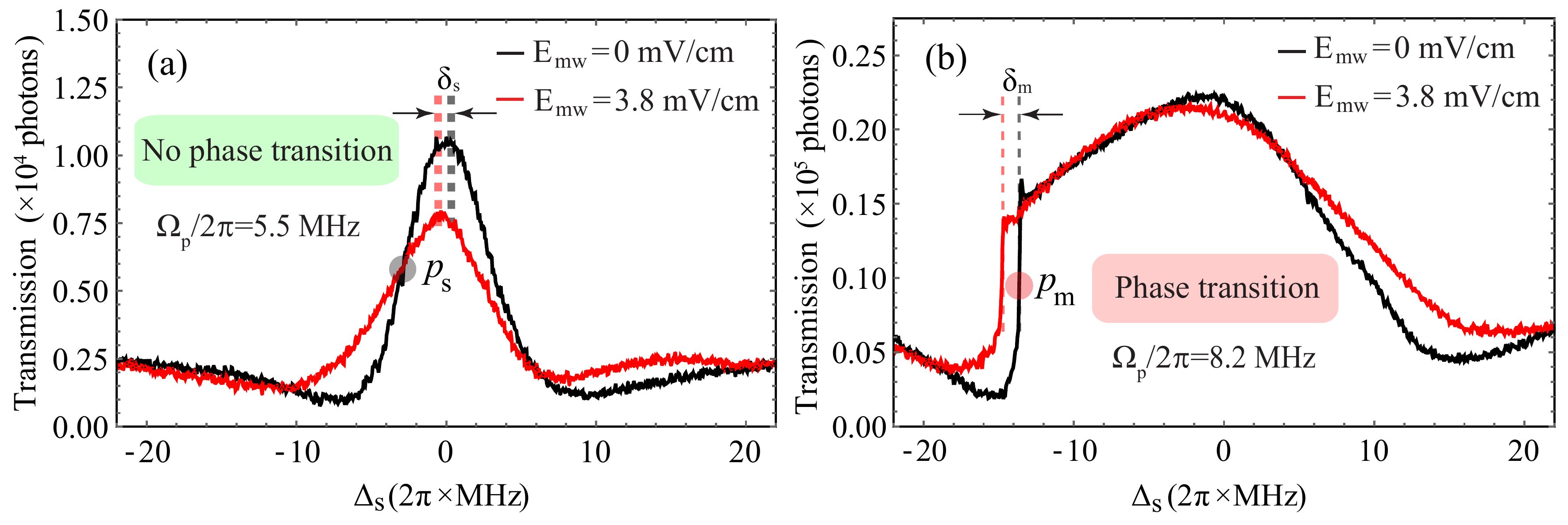}\caption{\textbf{Change of transmission spectra by application of MW fields.} (a) The transmission spectra under the field amplitude $E_{\mathrm{mw}}=0$ $\mathrm{mV/}\mathrm{cm}$ (black) and $E_{\mathrm{mw}}=3.8$ $\mathrm{mV/}\mathrm{cm}$ (red) with probe Rabi frequency $\Omega_{p}=2\pi\times5.5$ MHz below the critical value $\Omega_{p,c}=2$ for the phase transition. (b) The transmission spectra under the field amplitude $E_{\mathrm{mw}}=0$ $\mathrm{mV/}\mathrm{cm}$ (black) and $E_{\mathrm{mw}}=3.8$ $\mathrm{mV/}\mathrm{cm}$ (red) with the probe Rabi frequency $\Omega_{p}=2\pi\times8.2$ MHz, above the critical value for the phase transition. In these two cases, the frequency of the applied MW field is set as $2\pi\times16.68$ $\mathrm{GHz}$. The big circular points $p_{s}$ and $p_{m}$, marked with red and gray in Fig.~\ref{shifted spectra}(a) and (b), correspond to the position of the steepest slope. The direction of scanning $\Delta_{c}$ is from the red-detuning towards the blue detuning. Here, $\Delta_{s}$ is swept from -$2\pi\times$30 MHz to -$2\pi\times$24 MHz with a sweep rate $v_{s}$ = $2\pi\times$0.0055 MHz/$\mu$s.
}
\label{shifted spectra}
\end{figure*}

In the experiment, we sweep the detuning $\Delta_{s}$ and observe the transmission at each $\Delta_{s}$. As explained above, the FI is not only governed by the number of excited and thus interacting Rydberg atoms but also depends on the measurement time $t$ \citep{Theodoros2021Criticality}, defined as the time the probe laser explores a small interval around each detuning $\Delta_{s}$. In Fig.~\ref{Fishervstime}(a-c) we consider the behavior above criticality with $\Omega_{p}\text{/}2\pi=7.9$ MHz and in Fig.~\ref{Fishervstime}(d-f), we consider the behavior below criticality with $\Omega_{p}\text{/}2\pi=5.5$ MHz. We observe that for $\Omega_{p}\text{/}2\pi=7.9$ MHz, the transmission profile near the critical point becomes steeper as the measurement time $t$ is increased, while, for $\Omega_{p}\text{/}2\pi=5.5$ MHz, the transmission spectra are almost identical. This implies that the FI is inclined to be linearly dependent on the time $t$ for the data in Fig.~\ref{Fishervstime}(d-f)], while a different dependence appears for the data in Fig.~\ref{Fishervstime}(a-c)].

The values of the Fisher Information  are shown in Fig.~\ref{Fishervstime}(g) for different measurement times $t$, we find that the FI is fitted well by the form  $F=A(t/t_{0})^{\lambda}$, where $A=2.0\times10^{4}$ $\textrm{MHz\ensuremath{^{-2}}}$ and $\lambda=1.28$ for $\Omega_{p}\text{/}2\pi=7.9$ MHz, while $A=10$ $\textrm{MHz\ensuremath{^{-2}}}$ and $\lambda=1$  for $\Omega_{p}\text{/}2\pi=5.5$ MHz. In our system, when $t=1$ ms we achieve a large enhancement ratio up to $10^{3}$ by comparing these two cases. We conclude that one can extract more information by the interacting  many-body system than by  independent systems, and that we can extract even more information by continuous measurements for long times. The non-integer power-law dependent behavior of the fit to the FI is caused by the critical slowing down and thus deviation of the atomic dynamics from the stationary state around the critical point \cite{zurek2005dynamics,clark2016universal,keesling2019quantum}. This smooths the maximum slope and causes a more than linear suppression of the FI for the shorter measurement times in Fig.~\ref{Fishervstime}(a-c). We also plot the derivative $d\overline{\mu}/d\Delta_{s}$ against the detuning $\Delta_{s}$ in the vicinity of the critical detuning at $t=10\mu s$, see the inset in Fig.~\ref{Fishervstime}(b). We find that the derivative $d\overline{\mu}/d\Delta$ has a power law dependence on detuning $d\overline{\mu}/d\Delta_{s}=\chi\left|\Delta_{s}/\Delta_{0}+11.3\right|^{-\alpha}$, where $\chi=0.02$ $\textrm{MHz}^{-1}$, $\Delta_{0}/2\pi$ = 1 MHz, and $\alpha=2\pm0.1$ is the fitted power-law exponent. This detuning-dependent susceptibility is caused by the increased interaction near the critical point \citep{Trenkwalder_2016}, as the change of detuning tunes the Rydberg population $\rho_{rr}$ and hence the interaction. For the non-interacting case, the atomic system follows the stationary state even for the fast sweeps in our experiments, and the FI is linearly dependent on time $t$. There is more noise in Fig.~\ref{Fishervstime} (c) and (f) than Fig.~\ref{Fishervstime} (a) and (d), due to low frequency noises appearing for the longer measurement times. 

\begin{figure*}
\centering
\includegraphics[width=2\columnwidth]{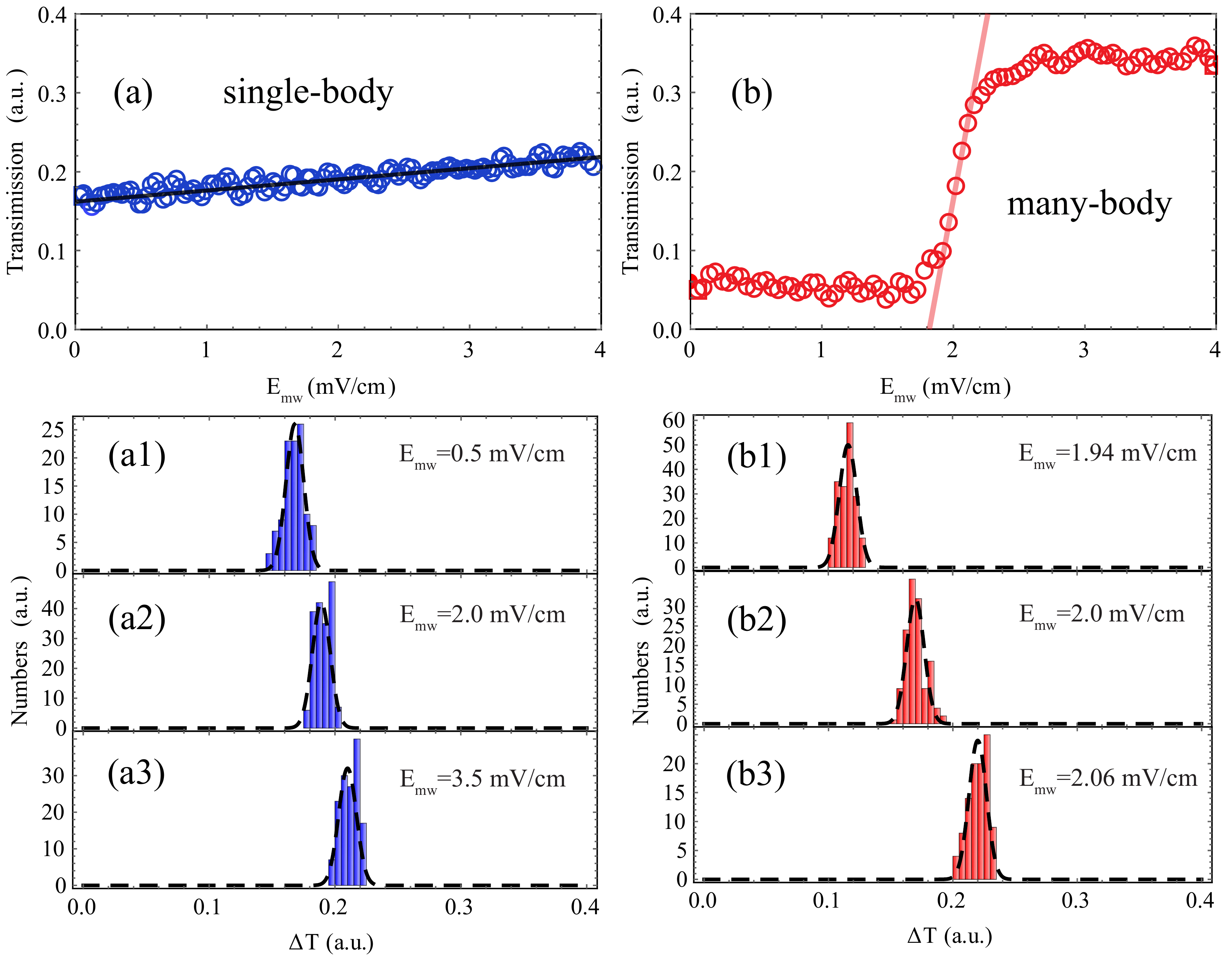}\\
\caption{\textbf{Transmission under different amplitudes of the MW field for the single-body case (a) and the many-body case (b).} The amplitude $E_{mw}$ is changed from $E_{mw}$ = 0 mV/cm to $E_{mw} $ = 4 mV/cm in steps of duration 5 $\mu$s. The black and red solid lines are the fit linear functions y=0.014 (x + 11.5) and y=0.89 (x - 1.81) respectively. (a1-a3) and (b1-b3) are the critical-point histogram of the transmission distribution under the different amplitudes of the MW field with the single-body case and the many-body case, respectively. The dashed lines are fitted Gaussian functions. The data 
are taken under equivalent experimental conditions (such as scan rate, acquisition time and averaging) in the few- and many-body cases}\label{response}
\end{figure*}

\textbf{Transmisssion spectra with and without MW fields.} The advantages for metrology appear due to the critical response of the system upon variation of external perturbations. To further study this critical response we apply a MW electric field with amplitude $E_{\mathrm{mw}}$ and detuning $\Delta_{\mathrm{mw}}$ to continuously drive the Rydberg transition $51D_{3/2} - 52P_{1/2}$. The main effect of the microwave field here is to (1) induce a small AC Stark shift which moves the critical point, (2) change the population of the Rydberg atoms while it has a negligible effect on the $C_{6}$ van der Waals coefficient and the resonant dipole-dipole interactions. As shown in Fig.~\ref{shifted spectra}(a) and (b), this shifts the transmission spectra of the Rydberg system subject to the probe Rabi frequencies $\Omega_{p}/2\pi=5.5$ MHz and $\Omega_{p}/2\pi=8.2$ MHz, corresponding to the non-interacting and interacting many-body systems, respectively.  When we apply the MW field with $E_{\mathrm{mw}}=3.8$ $\mathrm{mV/}\mathrm{cm}$, the spectra show a small red-shift $\delta_{s}$  [Fig.~\ref{shifted spectra}(a)] and $\delta_{
m}$ [Fig.~\ref{shifted spectra} (b)], respectively. For the many-body system, the high sensitivity of the frequency shift allows us to sense the strength $E_{mw}$ of an applied MW field. The FI \textbf{$F(\Delta)|_{\Delta_{s}=\Delta_{c}}$} for the black data in Fig.~\ref{shifted spectra} (a) is $F$ = $1.4\times 10^{-3} \textrm{MHz}^{-2}$, and is much smaller than that for the black data in Fig.~\ref{shifted spectra} (b) [$F$ = 0.27 $\textrm{MHz}^{-2}$]. The corresponding minimum uncertainty $\delta\Delta/2\pi \sim$ 0.3 MHz corresponds to an uncertainty of the applied field $\Delta E_{mw}$ = 1.9 mV/cm by considering the energy shift proportional to $E_{mw}^2$ when $E_{mw}$ is small [here the stark shift $\delta_{m}\sim E_{mw}^2$ follows by a Taylor expansion, $\delta_{{\rm m}}\sim-\Delta_{mw}/2+\sqrt{\Delta^{2}_{mw}+\Omega^{2}_{mw}}/2$], see more data in the supplementary materials. As a result, the many-body system can sense the strength $E_{mw}$ = 3.8 mV/cm directly by measuring the spectrum shift [$\delta_{m}$ = 2$\pi\times1.2(0.3)$ MHz]. In comparison, the spectral shift $\delta_{s}$ is  indistinguishable for independent atoms, which are thus not sensitive enough to sense the same MW field by monitoring of the spectral shift.

\textbf{Optical response under electric fields with different amplitudes.} The many-body metrological ruler has a thinner tick mark and thus a better precision than the single-body ruler, because the optical response is stronger in the many-body case when subject to a small frequency shift. We can also measure the optical transmission at the position of the steepest slope $p_{w1}$ (Fig.~\ref{setup}(b)) when the atoms are subject to a MW electric field $E_{mw}\textrm{sin}(f_{0}t)$ where $f_{0}/2\pi$ = $16.60$ GHz is near resonant with the RF transition $51D_{3/2}$  - $52P_{1/2}$. 

We measure the transmission when increasing the amplitude of the MW with the detuning $\Delta_s$ fixed  near the maximum slope under the many- and single-body conditions, as shown in Fig.~\ref{response} (a) and Fig.~\ref{response} (b), respectively. For the single-body condition, the transmission is not sensitive to the variance of the amplitude $E_{mw}$. For the many-body condition, the change of the $E_{mw}$ makes the system cross the critical point and the transmission signal is highly sensitive to the field around values of  $E_{mw}=2$ mV/cm (this position can be tuned by the coupling detuning), as shown in Fig.~\ref{response} (b). To evaluate the sensitivity, we fit the data near criticality with a linear function $y=k(x+x_{0})$ ($k=\overline{\mu}'(E_{mw})$), and obtain the ratio of the slopes $k_2/k_1=63.57$, where $k_1$ and $k_2$ represent the linear coefficients for the single- and many-body cases. As the almost same variance $Var(\mu)$ for these two cases, see the following figures (a1-a3) and (b1-b3), we can obtain enhanced ratio for the FI: ($k_{2}^2/Var(\mu))/(k_{1}^2/Var(\mu))>4000$. From the $E_{mw}$-dependent transmission, we can distinguish the standard deviation of the amplitude $\delta E_{mw}$ = 1.4 mV/cm for the non-interacting case and $\delta E_{mw}$ = 22 $\mu$V/cm for the interacting many-body case with 5 $\mu s$ data acquisition time per data point. By considering multiple sequential independent measurements, we estimate the equivalent sensitivity 49 nV/cm/$\textrm{Hz}^{1/2}$. 

\section*{Discussion}

Although the previous work in  Ref.~\citep{wade2018terahertz} clearly shows effects sensitive to the electric field,  that work mainly elaborated on how the presence of ionized Rydberg atoms induce a linear shift of the critical point. In contrast, the critical behavior of the interacting many-body system  has not previously been employed for sensing. The criticality induced by the interacting Rydberg atoms depends on the Rydberg atom number $N$ and  interaction strength $V$ \cite{ding2019Phase}, and the increase of the population $\rho_{rr}$ or the interaction strength $V\rho_{rr}$ enhances the non-linearity of the criticality. In our system, we have a large interacting number of atoms, and the energy splitting is far from the one of $N$ isolated systems,
\iffalse
. Also in this respect, our experiment thus presents experimental proof of principle of advantages of using the criticality of interacting systems in many-body metrology, 
\fi
cf. previous theoretical works \citep{macieszczak2016dynamical, PhysRevA.96.013817,raghunandan2018high,PhysRevLett.124.120504, PhysRevLett.126.010502,PhysRevLett.126.200501}. Specifically, the interaction induced non-linear FI dependence on measurement time of a single frequency scan shows an unique advantage on sensing, which agrees with the theoretical simulations in the supplemental materials. 

In summary, we have demonstrated the critical behavior of interacting Rydberg atoms and characterised its metrological consequences. The Fisher information for the estimation of a weak microwave field shows an enhancement of order $10^3$ by the use of interacting many-body systems. Concerning the use of the Fisher Information and Cram\'er-Rao bound, we note that not only the narrow detuning interval with the highest slope, but the entire transmission signal contributes in an integral manner to the sensitivity of the experiments, {\it cf.} a similar analysis of spatial image processing \citep{Negretti_2008,delaubert2008quantum}. Our analysis captures the main contribution to that integral and thus constitutes a lower limit to the FI. Passing from the measurement of frequencies, the experiments derive their improved sensitivity towards MW electric fields and show that the Rydberg non-equilibrium system can act as a versatile high-sensitivity metrological resource.

\section*{Method}

\textbf{Experiment setup.} 
We adopt a two-photon transition scheme to excite an atomic ground-state to a Rydberg-state, using a probe field resonantly driving the atomic transition $\ensuremath{5S_{1/2},F=2\rightarrow5P_{1/2},F'=3}$, and a coupling field, driving the transition $\ensuremath{\ensuremath{5P_{1/2},F'=3\rightarrow51D_{3/2}}}$. An MW electric field 1 (or 2) may be applied to drive an RF transition between two different Rydberg states $51D_{3/2}$ and $52P_{1/2}$(or $50F_{5/2}$). The MW electric fields used in our experiment are generated by two RF sources and two frequency horns. A 795 nm laser is split by a beam displacer into a probe beam and an identical reference beam, which are both propagating in parallel through a heated Rb cell (length $10\,\mathrm{cm}$). The temperature is set as $44.6^{\circ}\mathrm{C}$, corresponding to the atomic density of $9.0\times10^{10}$ $\mathrm{cm^{-3}}$. One probe beam is overlapped with a counter-propagating coupling beam to constitute the Rydberg-EIT process. The two transmission signals are detected on a differencing photodetector.

\textbf{Generation and calibration of MW fields.}
The MW fields used in our experiment are generated by two RF sources and two frequency horns. The RF source 1 works in the range from DC to 40 GHz, another is in the range from DC to 20 GHz. The frequency horns are set close to the Rb cell. The RF frequency between Rydberg D and P/F states are calculated according to the algorithm in Ref.~\citep{vsibalic2017arc}. We use a spectrum analyzer (Ceyear 4024F, 9 kHz $ \sim$ 32 GHz) and an antenna (380 MHz $\sim$ 20 GHz) to receive the MW fields then to calibrate the amplitude of MW fields in the centre of the Rb cell.

\textbf{Fisher information and Cram\'{e}r-Rao bound.}
In parameter estimation, the Cram\'{e}r-Rao bound sets a lower limit to the statistical estimation error by $\nu$ independent experiments, $(\delta  \theta)_{min} = 1/\sqrt{\nu\cdot F(\theta)}$. Here $F(\theta)$  is the Fisher Information (FI) which has the value

\begin{equation}
F\text{(\ensuremath{\theta})}
=\sum_{\mu}\frac{1}{L(\mu,\theta)}\left(\frac{\partial L(\mu,\theta)}{\partial\theta}\right)^{2}, 
\iffalse
-\sum_{\mu}L(\mu,\theta)\frac{\partial^{2}Log[L(\mu,\theta)]}{\partial\theta^{2}},
\fi
\end{equation}
where $L(\mu, \theta)$ is the likelihood function for the possible measurement outcome $\mu $, conditioned on the parameter $\theta$ \citep{pe1998}.
\iffalse
The FI is the average of $-\partial^{2}Log[L(\mu,\theta)]/\partial\theta^{2}$, weighted with the outcome distribution, and it thus quantifies how much a given outcome in an experiment reveals, on average, about the unknown parameter $\theta$. 
\fi
In our experiment we subtract two counting signals which may both be well described by Poisson distributions with mean values $\overline{\mu}_1$ and $\overline{\mu}_2$, and the same values for their variances. Both count numbers are large, and in our experiments the observed noise is dominated by electronic noise, and hence the distributions are both well approximated by Gaussian distributions. The difference signal is thus described by a Gaussian distribution with mean value $\overline{\mu} = \overline{\mu}_1 - \overline{\mu}_2$ and variance $\sigma^2 = \overline{\mu}_1 + \overline{\mu}_2$. For a Gaussian distribution with the likelihood \citep{mardia1984maximum,miller1974complex}
\begin{equation}
L(\mu,\theta)=\frac{1}{\sqrt{2\pi\sigma^2}}e^{-\frac{1}{2}(\frac{\mu-\overline{\mu}(\theta)}{\sigma})^{2}}
\end{equation}
one can see that
\begin{equation}
\frac{\partial L(\mu,\theta)}{\partial\theta}=\left(\frac{\mu-\overline{\mu}}{\sigma^{2}}\right)L(\mu,\theta)\overline{\mu}'(\theta).
\end{equation}
We hence obtain

\begin{align}
F\text{(\ensuremath{\theta})}
&=\sum_{\mu}\frac{1}{L(\mu,\theta)}\left(\left(\frac{\mu-\overline{\mu}}{\sigma^{2}}\right)L(\mu,\theta)\overline{\mu}'(\theta)\right)^{2}\nonumber \\
  &=\sum_{\mu}L(\mu,\theta)\frac{(\mu-\overline{\mu})^{2}}{\sigma^{4}}\left(\overline{\mu}'(\theta)\right)^{2}\nonumber \\
  &=\frac{Var(\mu)}{\sigma^{4}}\left(\overline{\mu}'(\theta)\right)^{2}=\frac{\left(\overline{\mu}'(\theta)\right)^{2}}{\sigma^{2}}
\end{align}
where $\overline{\mu}'(\theta)$ denotes the derivative of the mean $\overline{\mu}(\theta)$ w.r.t $\theta$.  There is a further contribution to the FI due to the dependence of the variance $\sigma^2$ on $\theta$. Its value is $\frac{1}{2\sigma^4} (\frac{d(\sigma^2)}{d\theta})^2$, and for our system it plays a less important role.

\textbf{Signal to ratio analysis.} 
For a given Gaussian incident probe and reference beams with mean photon signal $\mu_0$ in one second, the on-resonance absorption coefficient of the atoms and the loss of the optical path is $1-\zeta$ (corresponding to the transmission ratio $\zeta$), a differencing photodetector with efficiency $\eta$, records a Poisson distributed number of clicks with mean value $\eta \mu_0$ and variance $\eta \mu_0$ per second, which outputs a voltage signal. The difference between two beams is much weaker and has the mean value
\begin{equation}
\mu=\zeta\eta\mu_{0}[1+\varepsilon(\Delta_{s})]t-\zeta\eta\mu_{0}t=\zeta\eta\mu_{0}\varepsilon(\Delta_{s})t
\end{equation}
where $\varepsilon(\Delta_{s})$ is the transmission probability induced by Rydberg-EIT effect, $t$ is the considered time interval. As the input photon number of each beam is very large $\mu_{0}\sim 10^{14}$ photons per second for $\Omega_{p}/2\pi$ = 7.9 MHz, the variance of the difference of the two beams per second is the sum of the means because the difference of two Gaussian-distributed variables is also a Gaussian distributed variable: $Var(\mu)$ = $2\zeta\eta\mu_{0}+\zeta\eta\mu_{0}\varepsilon(\Delta_{s})$ $\sim$ 2$\zeta\eta\mu_{0}$. $\zeta= 20.6\%$ for $\Omega_{p}/2\pi$ = 7.9 MHz, $\zeta= 14.6\%$ for $\Omega_{p}/2\pi$ = 6.5 MHz, $\zeta= 8.7\%$ for $\Omega_{p}/2\pi$ = 5.5 MHz. 
The output voltage signal of the differencing photodetector could be converted into the photon number by a voltage conversion ratio $G=5.3\times10^{7}$ V/W. The 0.5 voltage output signal corresponds to $\sim1\times10^{10}$ photon numbers per second. The coupling detuning $\Delta_{s}$ is swept with a rate of $v_{s}$ and with M sampling rate [means that there are average M data points in the swept detuning $\Delta_{s}$], the transmission spectrum could be measured by accumulating the photon numbers in each detuning interval. The fast scan accumulate small photon numbers for each interval, while the slow scan get large photon numbers. If we scan the detuning of coupling laser $\Delta_{s}$ from red- to blue-detuning and vice versa, we could observe the bistability, see also in Refs.~\citep{carr2013nonequilibrium,weller2016charge,ding2019Phase}. The bistability shifted by the MW fields could be found in the supplementary materials.

\begin{figure}
  \centering
  \includegraphics[width=1\columnwidth]{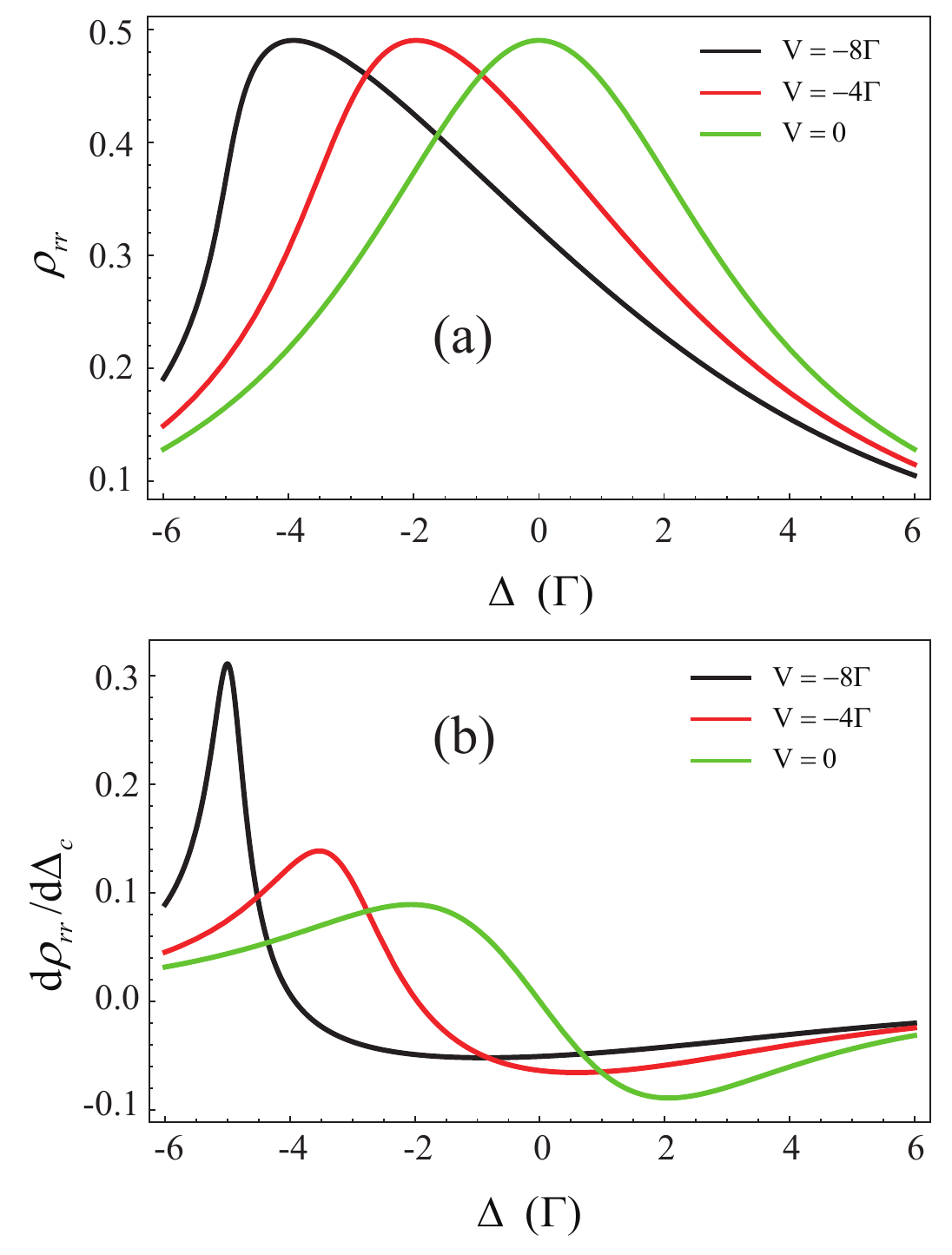}\\
  \caption{\textbf{Theoretical simulations of interacting two-level atoms.} (a) Rydberg state population $\rho_{rr}$ as a function of laser detuning $\Delta$ for different interaction strengths $V$.(b) The derivative of Rydberg population with respect to $\Delta$. }\label{direvative}
\end{figure}

\textbf{Non-linearity of interacting atoms.} 
To elucidate the sensitivity of the Rydberg atoms to the frequency for different interaction strength, we consider a two-level atoms model with ground state $\left| g \right\rangle$ and Rydberg state $\left| r \right\rangle$ (with spontaneous radiation rate $\Gamma$), which are coupled by a laser with Rabi frequency $\Omega$ and detuning from resonance $\Delta$. After mean-field approximation (i.e., $\Delta \rightarrow \Delta - V\rho_{rr}$, where $V$ is the many-body interaction term from dipole interaction or ions collisions, and $\rho_{rr}$ is the population of the Rydberg state), the steady-state solution for a two-level optical Bloch equation is \cite{lee2012collective}
\begin{equation}
\begin{array}{cc}
\dot{\rho}_{\mathrm{gr}}=i\frac{\Omega}{2}\left(\rho_{\mathrm{rr}}-\rho_{\mathrm{gg}}\right)+i\Delta_{\mathsf{\mathrm{eff}}}\rho_{\mathrm{gr}}-\frac{\Gamma}{2}\rho_{\mathrm{gr}}\\
\dot{\rho}_{\mathrm{rr}}=-i\Omega\left(\rho_{\mathrm{gr}}-\rho_{\mathrm{rg}}\right)-\Gamma\rho_{\mathrm{rr}}
\end{array}
\end{equation}
where $\Delta_{\mathrm{eff}}$ is the effective detuning $\Delta_{\mathrm{eff}}=\Delta - V\rho_{rr}$ by considering an interaction strength $V$. We obtain an equation about $\rho_{rr}$ for bistability as follows
\begin{equation}
V^{2} \rho_{\mathrm{rr}}^{3}-2 V \Delta \rho_{\mathrm{rr}}^{2}+\left(\Delta^{2}+\Omega^{2} / 2+\Gamma^{2} / 4\right) \rho_{\mathrm{rr}}-\Omega^{2} / 4=0
\end{equation}

By taking the derivative of both sides, we finally get the relationship between the slope of the steep edge of the hysteresis loop $\max(d\rho_{rr}/d\Delta) $ and the interaction of Rydberg atoms $V$.
\begin{equation}\label{eq:derivative}
  \frac{\mathrm{d}\rho_{rr}}{\mathrm{d}\Delta}  = -\frac{8 \left(\Delta  \rho_{rr} -\rho_{rr} ^2 V\right)}{\Gamma ^2+4 \Delta^2+12 \rho_{rr} ^2
   V^2-16 \Delta  \rho_{rr}  V+2 \Omega ^2}
\end{equation}
The critical point is defined when this derivative reaches infinity $\mathrm{d}\rho_{rr}/\mathrm{d}\Delta$ $\Rightarrow$ $\infty$, from which the threshold of the Rydberg population is obtained, i.e., 
\begin{equation}\label{eq:threshold}
\rho_{th} = \frac{\sqrt{-3 \Gamma ^2 V^2+4 \Delta ^2 V^2-6 V^2 \Omega ^2}+4 \Delta  V}{6 V^2}. 
\end{equation}

The relation between $\rho_{rr}$, $ \mathrm{d}\rho_{rr}/\mathrm{d}\Delta$, and $\Delta$ is demonstrated in Fig.~\ref{direvative}. And by letting the derivative equals to 0, i.e., $ \mathrm{d}\rho_{rr}/\mathrm{d}\Delta = 0$, we obtain the analytical expression of maximum derivative $ \left.\mathrm{d}\rho_{rr}/\mathrm{d}\Delta\right|_{\Delta=\Delta_{c}}$ and the corresponding detuning $\Delta_{c}$ as follows
\begin{equation}\label{eq:Delta}
 \Delta_{c} = \frac{1}{6} \left(6 \rho_{rr}  V-\sqrt{3} \sqrt{\Gamma ^2+2 \Omega ^2}\right)
\end{equation}
\begin{equation}
\frac{\mathrm{d}\rho_{rr}}{\mathrm{d}\Delta}(\Delta_{c})=\frac{1}{V+\sqrt{(\Gamma^{2}+2\Omega^{2})/3\rho_{rr}^{2}}}
\end{equation}

\textbf{Data availability.} The data that support this study are available at Github \citep{zongkai_2022_6333027} (\href{https://github.com/ZongkaiLiu/many-body-enhanced-metrology}{many-body-enhanced-metrology}).

\bibliography{ref}

\textbf{Acknowledgment.} D-S.D thanks for discussions with professor Jun Ye from JILA. Z-K.L. appreciates instructive discussions with Dr. Tian-Yu Xie. We acknowledge funding from National Key Research and Development Program of China (2017YFA0304800), the National Natural Science Foundation of China (Grant Nos. U20A20218, 61525504, 61722510, 61435011), the Innovation Program for Quantum Science and Technology (2021ZD0301100), Anhui Initiative in Quantum Information Technologies (AHY020200), the Youth Innovation Promotion Association of Chinese Academy of Sciences under Grant No. 2018490, EPSRC through grant agreements EP/M014398/1 (``Rydberg soft matter''), EP/R002061/1
(``Atom-based Quantum Photonics''), EP/L023024/1 (``Cooperative quantum optics in dense thermal vapours''), EP/P012000/1 (``Solid State Superatoms''), EP/R035482/1 (``Optical Clock Arrays for Quantum Metrology''), EP/S015973/1 (``Microwave and Terahertz Field Sensing and Imaging using Rydberg Atoms''), the Danish National Research Foundation through the Center of Excellence for Complex Quantum Systems (Grant agreement No. DNRF156), as well as, DSTL, and Durham University.).

\textbf{Author contributions.} D-S.D. conceived the idea and implemented the physical experiments with Z-K.L. Z-K.L., D-S.D., and K.M. employ the Fisher information. D-S.D., Z-K.L., and K.M. derived the equations, plotted figures, and wrote the manuscript. All authors contributed to discussions regarding the results and analysis contained in the manuscript. D-S.D., B-S.S.,G-C.G., K.M., and C-S.A. support this project.

\textbf{Competing interests}. The authors declare no competing interests.

\end{document}

\title{Supplementary materials of enhanced metrology at the critical point of a many-body Rydberg atomic system}

\author{Dong-Sheng Ding$^{1,2,3,\textcolor{blue}{\ast},\textcolor{blue}{\dag}}$}
\author{Zong-Kai Liu$^{1,2,3,\textcolor{blue}{\ast}}$}
\author{Bao-Sen Shi$^{1,2,3,\textcolor{blue}{\ddagger}}$}
\author{Guang-Can Guo$^{1,2,3}$}
\author{Klaus Mølmer$^{4,\textcolor{blue}{\mathsection}}$}
\author{Charles S. Adams$^{5,\textcolor{blue}{\mathparagraph}}$}

\affiliation{$^1$CAS Key Laboratory of Quantum Information, University of Science and Technology of China, Hefei 230026, China}
\affiliation{$^2$CAS Center for Excellence in Quantum Information and Quantum Physics, University of Science and Technology of China, Hefei 230026, China}
\affiliation{$^3$Hefei National Laboratory, University of Science and Technology of China, Hefei 230088, China}
\affiliation{$^4$Aarhus Institute of Advanced Studies, Aarhus University, H{\o}egh-Guldbergs
Gade 6B, DK-8000 Aarhus C, Denmark\\
Center for Complex Quantum Systems, Department of Physics and Astronomy,
Aarhus University, Ny Munkegade 120, DK-8000 Aarhus C, Denmark}
\affiliation{$^5$Department of Physics, Durham University, South Road, Durham DH1 3LE, United Kingdom}
\date{\today}

\symbolfootnote[1]{These authors contributed equally to this work.}
\symbolfootnote[2]{dds@ustc.edu.cn}
\symbolfootnote[3]{drshi@ustc.edu.cn}
\symbolfootnote[4]{moelmer@phys.au.dk}
\symbolfootnote[5]{c.s.adams@durham.ac.uk}

\maketitle
\renewcommand{\thefigure}{S1}
\begin{figure*}[t]
\includegraphics[width=1\columnwidth]{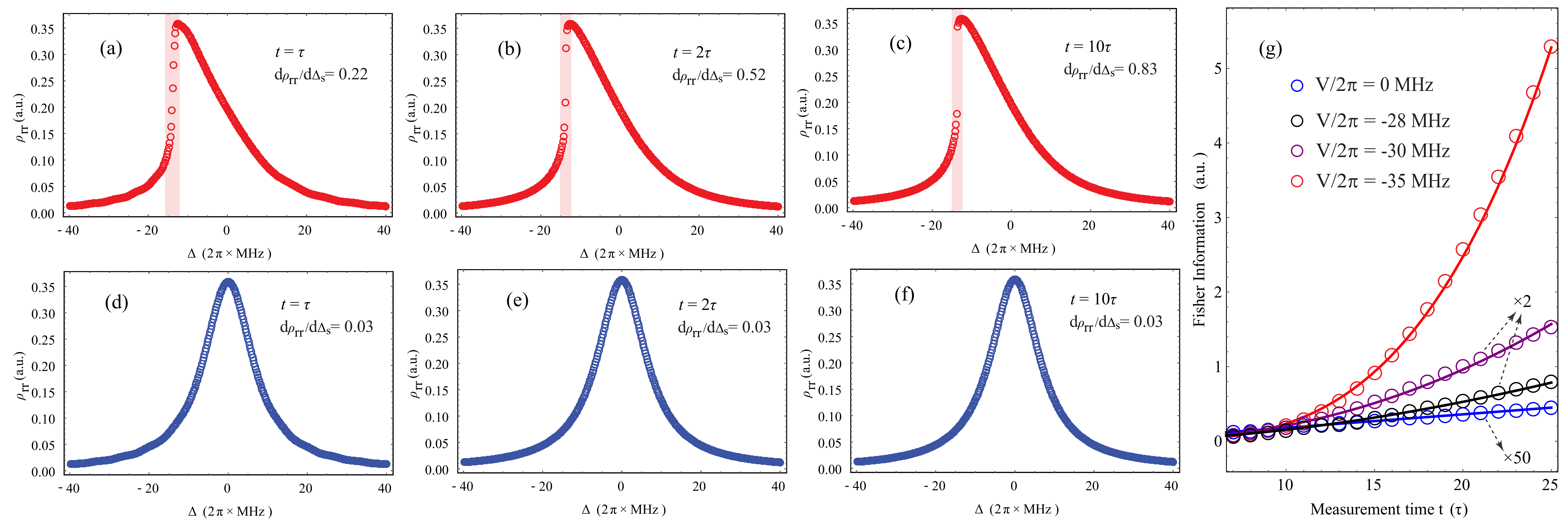}\caption{\textbf{Simulation on transmission dynamics of single- and many-body system.} The simulated population $\rho_{rr}$ against different sweep rates (or measurement time t) under the interaction strength $V/2\pi$=-35 MHz for (a-c) and $V$=0 MHz for (d-f).  In this simulation, $\Gamma/2\pi=8$ MHz and $\Omega/2\pi=9$ MHz. (g) The calculated Fisher information $\textbf{\ensuremath{F(\Delta)|_{\Delta=\Delta_{c}}=(\mathrm{d}\overline{\mu}/\mathrm{d\Delta})^{2}/\overline{\mu}}}$ under $V\text{/}2\pi=-35$ MHz (red), $V\text{/}2\pi=-30$ MHz (purple), $V\text{/}2\pi=-28$ MHz (black), and $V\text{/}2\pi=0$ MHz (blue), here $\overline{\mu}=\int\rho_{rr}dt$. The theoretical fit function for these two cases is $F=At^{\lambda}$, where $A=9.3\times 10^{-5}$ $\textrm{MHz\ensuremath{^{-2}}}$ and $\lambda=3.4$ for $V\text{/}2\pi=-35$ MHz, $A=1.3\times10^{-3}$ $\textrm{MHz\ensuremath{^{-2}}}$ and $\lambda=2.2$ for $V\text{/}2\pi=-30$ MHz, $A=2.6\times10^{-3}$ $\textrm{MHz\ensuremath{^{-2}}}$ and $\lambda=1.78$ for $V\text{/}2\pi=-28$ MHz, and $A=3.6\times10^{-4} \textrm{MHz\ensuremath{^{-2}}}$ and $\lambda=1$ for $V\text{/}2\pi=0$ MHz. The Fisher Information for the blue, black, and purple data are magnified manually by a factor 50, 2, and 2 respectively.}
\label{slope against time}
\end{figure*}

\textbf{Simulation on the time-dependent dynamics.} In order to analyze the dynamics at the criticality, we simulate the dynamics of the Rydberg population $\rho_{rr}$ under different measurement time $t$ by changing the sweep rate. By considering the time-dependent terms $\dot{\rho}_{\mathrm{gr}}$ and $\dot{\rho}_{\mathrm{rr}}$ in the equation (7) in main text, we solve the transient solution and obtain the Rydberg population $\rho_{rr}$ at different evolution time $t$ for each detuning $\Delta$, here the evolution time is equivalent to the time required for measurement. In this process, the sampling rates are 400, it means there are 400 data points in the average distribution from -$2\pi\times $40 MHz $\sim$ $2\pi\times $40  MHz. For the interaction $V/2\pi$ = -35 MHz, the derivative $\mathrm{d}\rho_{rr}\text{/\ensuremath{\mathrm{d}\Delta}}$ at the critical detuning decreases against the measurement time $t$, see the results in Figs.~\ref{slope against time}(a-c). For $t=\tau, 2\tau$ and $10\tau$, when $V/2\pi$ = -35 MHz, we obtain $\mathrm{d}\rho_{rr}\text{/\ensuremath{\mathrm{d}\Delta}}$ = 0.22, 0.52, and 0.83 respectively. While for $V$ = 0, the derivative $\mathrm{d}\rho_{rr}\text{/\ensuremath{\mathrm{d}\Delta}}$ = 0.03 at the steepest position, with no obvious change for different measurement time $t$. From these results, the derivative $\mathrm{d}\rho_{rr}\text{/\ensuremath{\mathrm{d}\Delta}}$ for the interacted many-body system has an obvious scale-up on the time $t$ by comparing that in the independent single-body system.

According to the mean field model, we have calculated the Fisher information $\textbf{\ensuremath{F(\Delta)|_{\Delta=\Delta_{c}}=(\mathrm{d}\overline{\mu}/\mathrm{d\Delta})^{2}/\sigma^{2}}}$ for the interacted many-body system [$V\text{/}2\pi=-35, -30, -28$ MHz] and independent single-body system [$V\text{/}2\pi=0$ MHz], which are given by the red and blue data of the Fig.~\ref{slope against time}(g). For the many-body case, there is a nonlinear dependence of the Fisher information with data acquisition time, while for the single-body case, the Fisher information increases linearly with time $t$. The nonlinear feature of the Fisher information is also dependent on the interaction strength $V$. The larger the interaction strength $V$, the larger is the exponent $\lambda$.

\renewcommand{\thefigure}{S2}
\begin{figure*}[t]
\includegraphics[width=2.1\columnwidth]{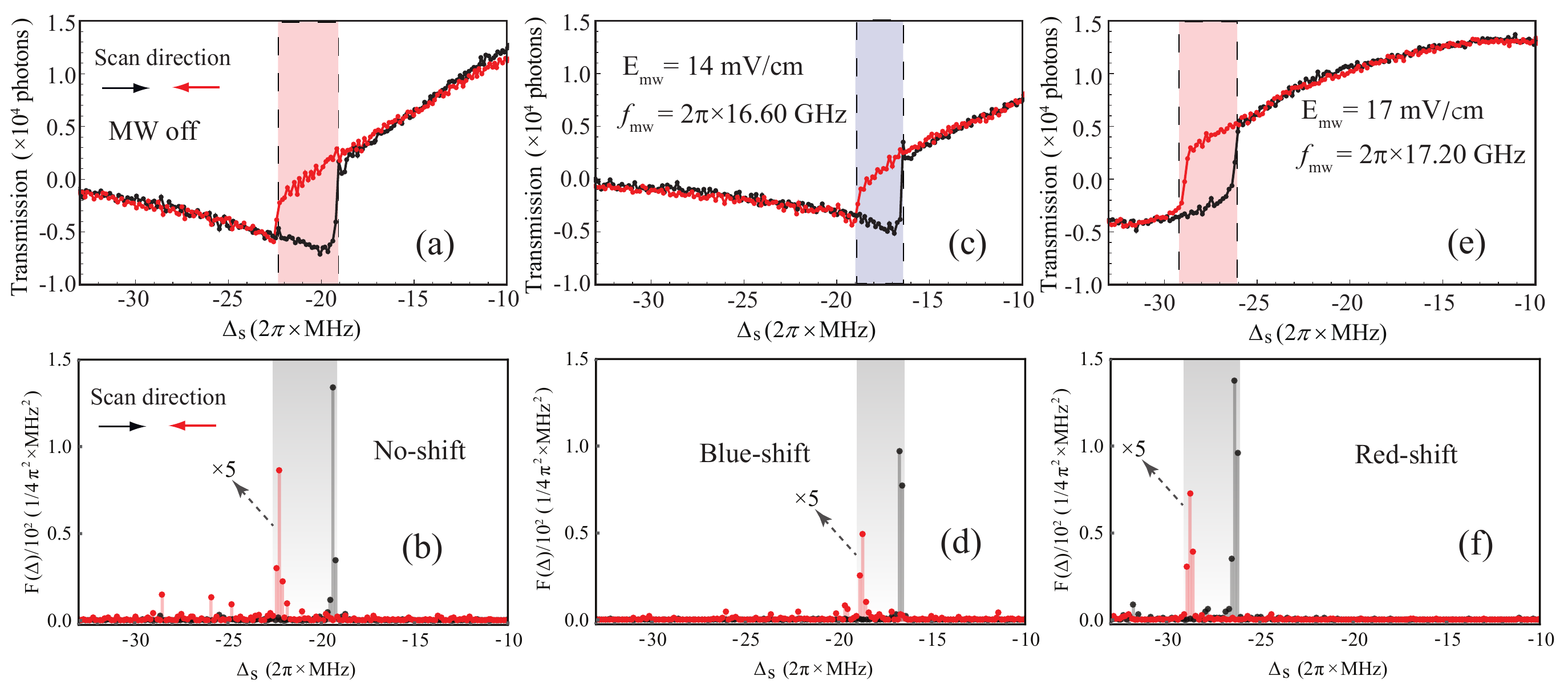}\caption{\textbf{Microwave electric fields shifted bistability}. The measured bistability without any MW electric field (a), with the MW electric field 1 (c), and with the MW electric field 2 (e). The bistability regions, as marked by the shaded areas in (c) and (e), show an obvious frequency shift by comparing with (a). In these cases, the intensity of the MW electric fields is set as $E_{\mathrm{mw1}}=14$
$\mathrm{mV/}\mathrm{cm}$ and $E_{\mathrm{mw}2}=17$ $\mathrm{mV/}\mathrm{cm}$. The corresponding Fisher information distributions with no-shift, blue-shift and, red-shift are given in (b), (d), and (f). The label $\times 5$ of red dots means that the Fisher information of the right-left scanning case is magnified manually by 5 times for better illustration. The shaded areas represent the width between the maximum Fisher information by scanning $\Delta_{s}$ with two different directions.}
\label{shift bistability}
\end{figure*}

\renewcommand{\thefigure}{S3}
\begin{figure*}[t]
\includegraphics[width=2\columnwidth]{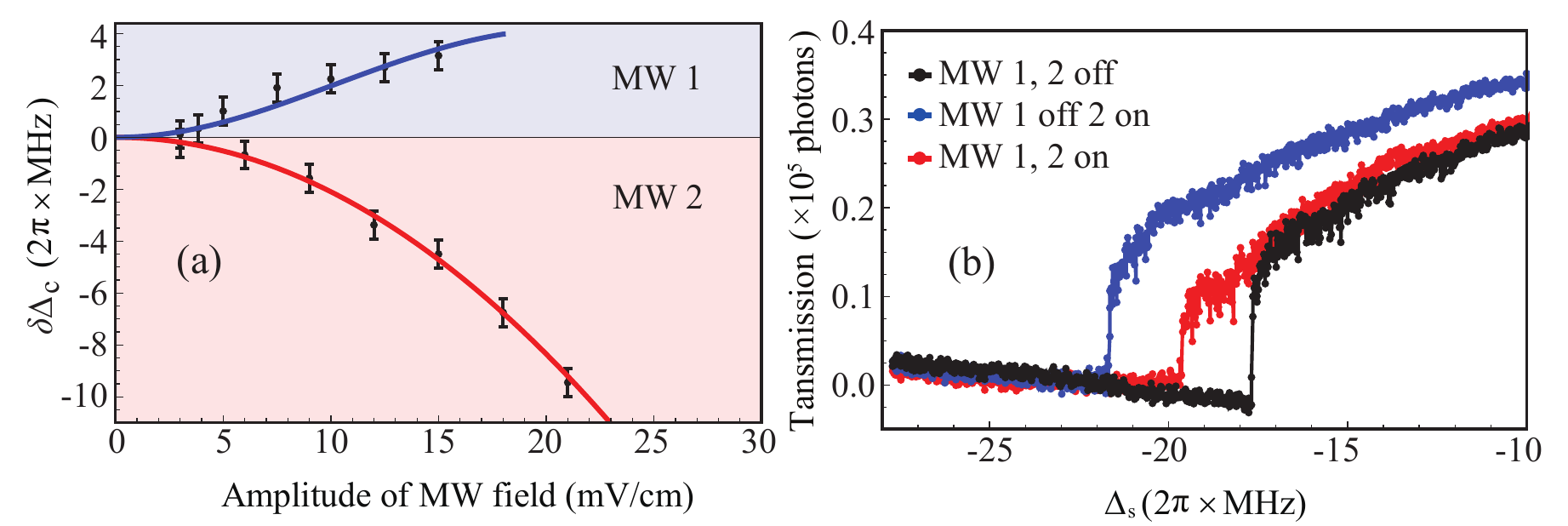}\caption{\textbf{Measurement of the shifted critical point}. (a) The measured frequency shifts at a transition point when coupled by the MW electric fields 1 and 2, respectively. The data are fit by $\Delta_{\mathrm{D}}$ with a MW electric field broadening $-(\Delta_{\mathrm{mw}}/2\pm\sqrt{\Delta_{\mathrm{mw}}^{2}+\Omega_{\mathrm{mw}}^{2}}/2)-\eta\Omega_{\mathrm{mw}}^{2}$
($\Omega_{\mathrm{mw}}=\mu_{\mathrm{RF}}E_{\mathrm{mw}}/\hbar$) with experimental parameters
(red) of $\Delta_{\mathrm{mw}}=-2\pi\times420$ MHz and a fitted broadening
coefficient $\eta=5.6\times10^{-4}$ $\mathrm{MHz^{-1}}$ and parameters (blue) of $\Delta_{\mathrm{mw}}=2\pi\times70$ MHz, $\eta=2.2\times10^{-3}$
$\mathrm{MHz^{-1}}$. Here $\mu_{\mathrm{RF}}=3320$ $ea_{0}$ is the dipole momentum of RF-transition $51D_{3/2}\rightarrow50F_{5/2}$ or $51D_{3/2}\rightarrow52P_{1/2}$,
$e$ is the electronic charge, and $a_{0}$ is the Bohr radius. (b) The measured phase transitions with the MW electric fields 1 and 2 both off (black), with the MW electric fields 1 off and 2 on (blue), and with the MW electric fields 1 and 2 both on (red). The frequencies of chosen RF transitions 1 and 2 are $f_{\mathrm{mw1}}=2\pi\times16.60$ $\mathrm{GHz}$ for $52P_{1/2}$ and $f_{\mathrm{mw2}}=2\pi\times17.22$ $\mathrm{GHz}$ for $50F_{5/2}$ respectively, and the intensities of the MW electric fields are set as $E_{\mathrm{mw}1}=7.8$ $\mathrm{mV/}\mathrm{cm}$
and $E_{\mathrm{mw2}}=9.0$ $\mathrm{mV/}\mathrm{cm}$.}
\label{shift bistability1}
\end{figure*}

\renewcommand{\thefigure}{S4}
\begin{figure*}
  \centering
  \includegraphics[width=2\columnwidth]{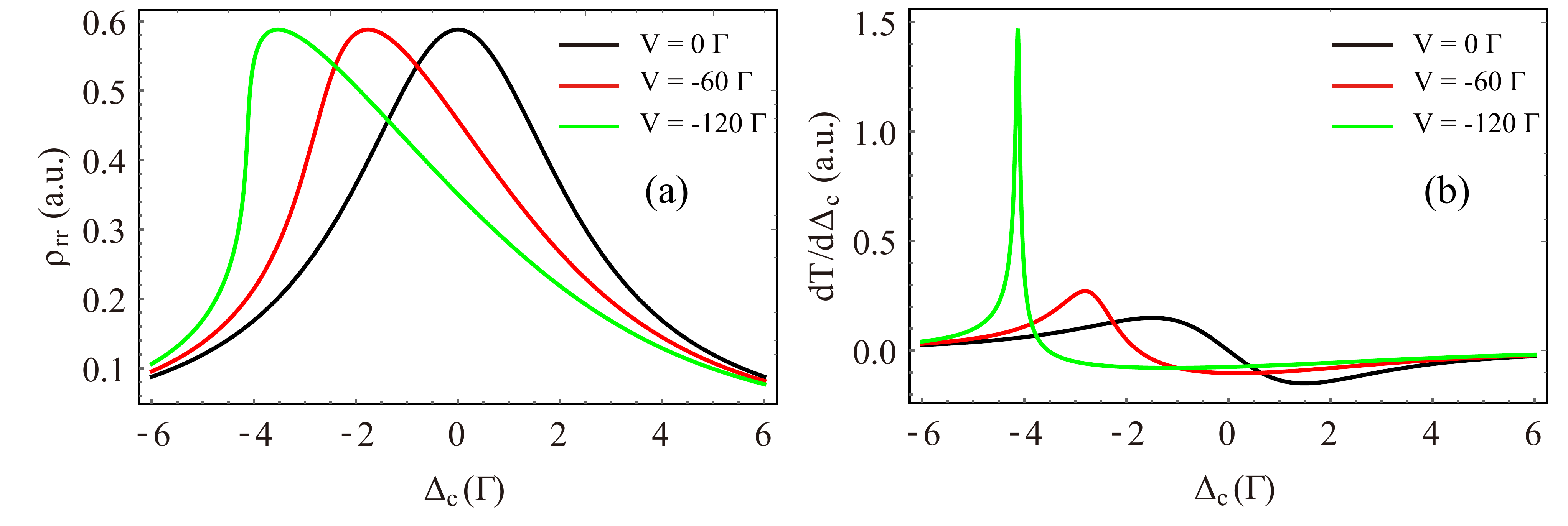}\\
  \caption{\textbf{Theoretical simulations of interacting three-level atoms.} (a) Rydberg state population $\rho_{rr}$ as a function of laser detuning $\Delta$ for different interaction strengths $V$.(b) The derivative of Rydberg population with respect to $\Delta$. }\label{three}
\end{figure*}

\textbf{Double MW shifted bistability.} In our experiment, the different scan directions give rise to hysteresis characterizing as the optical bistability. Due to the high-precision of the criticality on detuning, the bistability shift could be measured in small variance. We have studied the cases with two different MW electric fields $f_{\mathrm{MW1}}=2\pi\times16.60$ $\mathrm{GHz}$ and $f_{\mathrm{MW2}}=2\pi\times17.20$ $\mathrm{GHz}$ experimentally, both of these are under the red-detuning case. We first measure the bistability without the MW electric fields shown in Fig.~\ref{shift bistability}(a), and then open on the MW electric fields 1 and 2 individually, as given in Fig.~\ref{shift bistability}(c) and Fig.~\ref{shift bistability}(e). The bistability has a blue-shift of $2\pi\times3.7$ MHz for the MW electric field 1, and a red-shift of $-2\pi\times5.8$ MHz for the MW electric field 2. The reason why the different shift directions of the bistability for the states $52P_{1/2}$ and $50F_{5/2}$ is that the $52P_{1/2}$ state is lower (lambda-type energy diagram with $51D_{3/2}$ and $5P_{1/2}$) and the $50F_{5/2}$ state higher (ladder-type energy diagram with $51D_{3/2}$ and $5P_{1/2}$) in energy than the $51D_{3/2}$ state. The detuning $\Delta_{mw}$ is thus opposite for upper and lower Rydberg states. The double jumps in the hysteresis loop result in double FI peaks, as marked as red and black color in Fig.~\ref{shift bistability}(b), (d) and (f). The FI in the right-left scanning case [red in Fig.~\ref{shift bistability}(b), (d) and (f)] is smaller than that in the left-right scanning case [black in Fig.~\ref{shift bistability}(b), (d) and (f)]. This is because the bifurcation of many-body system at small detuning is more distinct than that at large detuning \citep{ding2019Phase}. For any given swept experiment, there is high-sensitivity criticality for frequency shift under $\Delta_{s}$ goes up [black in Fig.~\ref{shift bistability}(b), (d) and (f)] or drop down [red in Fig.~\ref{shift bistability}(b), (d) and (f)].

\textbf{Measurement of the shifted critical point.}
We measure the frequency shift of the transition point [or the maximum FI] for these two cases, as shown in Fig.~\ref{shift bistability1}(a). If tuning the shift with multiple MW electric fields, we could change the shift in back and forth by multiple MW electric fields. To demonstrate this effect, we record the transmission spectra with the MW electric fields 1 and 2 both off (black), the MW electric fields 1 off and 2 on (blue), and the MW electric fields 1 and 2 both on (red), as shown in Fig.~\ref{shift bistability1}(b). Without the MW electric fields, the phase transition appears at $-2\pi\times20.2$ $\mathrm{MHz}$. If applying the MW electric field 2, the phase transition point is shifted to $-2\pi\times22.6$ $\mathrm{MHz}$, but shifted back to $-2\pi\times21.4$ $\mathrm{MHz}$ when the MW electric field 1 is on. These show that the criticality of phase transition is red-shifted by the MW electric field 2, and then could be shifted back a little bit due to the opposite shifts for $52P_{1/2}$ and $50F_{5/2}$. When increasing $E_{\mathrm{mw}}$, the phase transition is inclined to disappear because the Rydberg populations are decreased below the driving of the MW electric fields as explained above.

\textbf{Three level model with mean field approximation.} In our case, the model considered in the simulation is two level model which is slightly different from that in experiment. However, the phenomena uncovered by  these two are the same. In Fig.~\ref{three}, the population of Rydberg atoms with different interaction strength $V$ varying with detuning of coupling light $\Delta_c$ and the corresponding derivative are also illustrated.

The three level model with mean field approximation is given in the following. The interaction of the
EIT light fields with an ensemble of three level atoms is governed
by the master equation for the single atom density matrix,
\begin{align}
\frac{{\mathrm{d}}}{{\mathrm{d}}t}\rho=-\frac{i}{\hbar}\left[H,\rho\right]+\frac{1}{\hbar}L[\rho]
\end{align}
where
\begin{align}
H=-\frac{\hbar}{2}\begin{pmatrix}0 & \Omega_{p} & 0\\
\Omega_{p} & -2\Delta_{p} & \Omega_{c}\\
0 & \Omega_{c} & -2(\Delta_{p}+\Delta_{c})
\end{pmatrix}
\end{align}
where $\Omega_{p(c)}$ and $\Delta_{p(c)}$ are the Rabi frequency and detuning of probe (coupling) light, respectively, and the Hilbert space is spanned by atomic ground state $\left| g \right\rangle$, excited intermediate state $\left| e \right\rangle$, and Rydberg state $\left| r \right\rangle$.

The Lindblad superoperator
\begin{equation}
\begin{array}{lc}
L[\rho]=\\
\frac{\hbar}{2}\begin{pmatrix}2\Gamma_{e}\rho_{ee} & -\Gamma_{e}\rho_{ge} & -\Gamma_{r}\rho_{gr}\\
-\Gamma_{e}\rho_{eg} & -2\Gamma_{e}\rho_{ee}+2\Gamma_{r}\rho_{rr} & -(\Gamma_{r}+\Gamma_{e})\rho_{er}\\
-\Gamma_{r}\rho_{rg} & -(\Gamma_{r}+\Gamma_{e})\rho_{re} & -2\Gamma_{r}\rho_{rr}
\end{pmatrix}
\end{array}
\end{equation}
where $\Gamma_{e}$ and $\Gamma_{r}$ accounts for the finite lifetime of $\lvert e\rangle$ and $\lvert r\rangle$.

In our mean field theory, the master equation is first solved in steady state, i.e., $\text{d}\rho/\text{d}t=0$. Then the detuning of the coupling light is modified with mean field approximation ($\Delta_c \rightarrow \Delta_c - V \rho_{rr}$) and the equations are iterated until convergence of $\rho_{rr}$.